\documentclass[12pt]{article}
  \usepackage{a4}      %
  \newcommand\ignore[1]{}
 \usepackage{epsfig,graphicx}
  
\begin{document}
  \hfill To be published in Proceedings of ECRYS-99,

\hfill Journ. de Physique, Coll., December 1999.

\vskip .6in

   {\Large{\bf{ Nonlinear conduction of sliding electronic crystals: 
Charge and Spin Density Waves.}}}
\medskip
 
 \vskip .1in
 {\centerline{ S. Brazovskii$^{1,3}$ and A. Larkin$^{2,3}$.}

 \vskip .1in

 {\it{ $^1$Laboratoire de Physique Th\'eorique et des Mod\`eles Statistiques, CNRS,}}

{\it{$^{\ }$ B\^at.100, Universit\'e Paris-Sud, 91406 Orsay, cedex, France}}

     {\it{$^2$Theoretical Physics Institute, Univ. of Minnesota, Minneapolis,
  MN 55455,}}

{\it{$^{\ }$ USA.}}

{\it{$^3$L.D.Landau Institute, Moscow, Russia.}}

\parskip 6pt

\vskip .2in
  \begin{narrow}
 {\bf Abstract.}  
A model  of  local metastable states (MSs)  due to the pinning induces plastic deformations  allows 
to describe the nonlinear $I-V$ curves in sliding density waves (DW). 
With increasing DW velocity $v$, the MSs of decreasing lifetimes $\tau \sim 1/v$  are accessed. 
The characteristic second threshold field  is reached when the shortest life time
configurations are accessed by the fast moving DW. 
Thus  the DW works as a  kind of a ``linear accelerator''  testing  virtual  states. 
\end{narrow}

{\bf 1. INTRODUCTION. LOCAL METASTABLE STATES.}
\vskip .1in

This talk summarizes a local pinning approach to time dependent properties of sliding superstructures,
especially on examples of Density Waves (DWs), see e.g. \cite{Proc89}.
We  have proposed \cite{Larkin94,Larkin95,Brazovskii96} a theory of metastable states (MSs)  
created by pinning-induced solitons or dislocation loops/lines (DLs) and have applied it to  
 the two commonly observed  features of DWs :

\noindent i. Low temperature $T$, low frequency $\omega$ anomalies of the dielectric susceptibility \\ 
 $ \varepsilon (T,\omega)$ \cite{Lasjaunias94};\\
ii. The $I-V$ curve with a  second threshold field ${\cal E}_2$ in
sliding regime {\cite{Mihaly88,Itkis90}}. (See \\
 more references in \cite{Brazovskii96}.)

The two types of pinning are usually distinguished \cite{Larkin79,FLR,Lee79}

{\em Collective or weak} pinning comes from an elastic interference of many
impurities. It originates the first threshold field ${\cal E}_1$ to initiate the
sliding. Its characteristic features are: low critical field for the friction at
rest ${\cal E}_1\propto 1/\varepsilon $; high  response $\varepsilon $,
correlation volumes and barriers $E_{bar}$ between MSs; huge relaxation
times $\tau \sim \exp [E_{bar}/T]$.
In DWs the collective pinning is  affected by the anomalous  Coulomb hardening limited only by  screening 
via  thermally activated normal carriers; it leads to the low $T$ release
of the collective pinning.

{\em Local or strong}  pinning comes from MSs at isolated  centers which provide
finite barriers, hence the reachable relaxation times. Except for low  $\omega $ or velocities $v$ 
when the collective creep can prevail, it becomes the main source of dispersion,
relaxation and dissipation.

Following  \cite{Larkin94,Larkin95,Brazovskii96} we recall now the origin of local MSs of the DW at a single
 impurity, their relations to solitons and dislocations and the time
 $t$ evolution in the course of the DW sliding.
Consider an isolated impurity at some point ${\bf r}_{i}$ interacting
with the DW $\sim \cos (\vec Q \vec r +\varphi)$.
 We introduce the positionally random phase $\theta =-\varphi_{0}-{\bf Qr}_{i}$ and the
DW phase at the impurity site $\psi =\varphi ({\bf r}_{i})-\varphi _{0}$,
both with respect to the reference value $\varphi _{0}$ within the large
correlated volume of the collective pinning. The averaging over
$\theta $ covers both  the distribution of 
${\bf r}_{i}$ in statics
and the motion of the DW in sliding when $\varphi_0=vt$.

For a weak impurity potential V, the local phase stays close to the one of the
correlated volume, $\vert \psi \vert \ll 1$. The state is unique: following one period $2\pi$
of $\theta $ the system returns to the same state of the same energy and  the averaged force  is zero. 
These impurities do not produce a local pinning proportional to their concentration (linear) $ n_i$ 
while their contribution $\sim n_i^2$ to the weak collective
pinning  is finite, given by the mean squared force.
Oppositely for large  $V$ the phase $\psi $ follows closely the impurity
position $\theta $ while allowing  to skip the quanta $\pm 2\pi $.
Apparently for $2\pi >\psi \approx \theta >\pi $ the state with 
$\psi ^{\prime }\approx \theta -2\pi ,\;\;0>\psi ^{\prime }>-\pi$
has a lower energy, the phase at the impurity being closer to the one in the bulk. 
Then such a site becomes bistable provided that the upper term
preserves its metastability. It really happens at large enough pinning
potentials $V>V_1$, but the bistability is maintained only at limited
intervals of $\theta $ which are terminated by  end points $\theta_e$. 
The two terms $E_{\pm}(\theta)$ cross at $\theta =\pi$
being separated by a barrier $E_{bar}$ which height (above the MS term) is maximal
$E_{bar}=E_\pi$ at $\theta =\pi$ and disappears at $\theta _e$.
At even higher $V>V_2$ there are no more end points. The MS term  accents without a terminations over the whole
 period of $\theta$, thus the whole $\psi=2\pi$ is accumulated at the impurity  in compare to the bulk. 
It results in  creation of  a diverging pair of $\pm 2\pi$  solitons. 
A curious effect of long range interactions between solitons is that 
 at large distances  the pair looses stability
and falls to the infinite separation at some critical $\theta$ before the circle is completed. 
The energy terms are shown at Fig.1.   Specifically for CDWs the characteristic length $l_s$, 
energy $E_s$ of the MS or of the single solitons are related to the transition temperature 
$T_c$ as $E_s\sim T_c, \ l_s\sim \hbar v_F/T_c$. 

\begin{figure}[thb]
\includegraphics{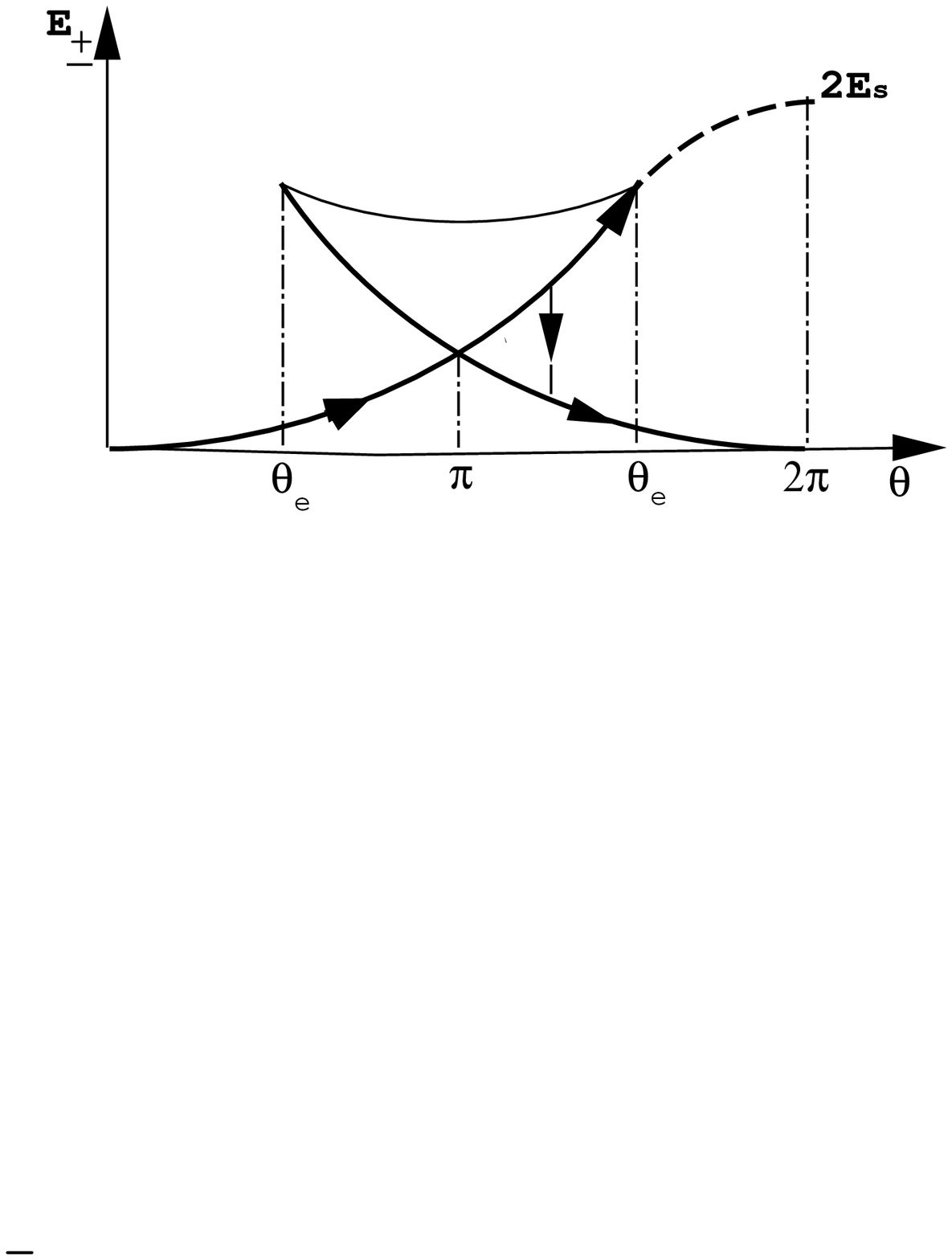}
\epsfxsize 1.5\hsize
\caption
{The evolution -arrows, of (meta) stable terms   and of 
  the barrier - thin line.  Dashed line: a case without end points.}
\end{figure}

Notice that  we can consider solitons as  nucleus dislocation loops which allows to generalize the DW language 
to other sliding structures. To demonstrate an universality of the MS
mechanism, let us consider a sliding 
Wigner or vortex crystal with strong attractive interaction with an impurity. 
In the moving frame the impurity passes along the raw of sites carrying the captured ``atom'' with it. 
Apparently during the second half of the full period it is profitable  to release the  atom to its position behind
 the impurity and to trap instead the atom from the approaching site which became closer. 
This switch to the lower term requires to pass over the barrier which is the origin of the MS. 
If the relaxation does not happen, the atom is dragged over the whole period to a new regular position 
which is equivalent to creation of a pair of DLs.

The relaxation of MSs is generally provided by their activation over
the barrier via structural fluctuations. 
But for CDWs there is also an external mechanism of relaxation by trapping a thermally activated
normal carrier. This process is facilitated by a double wall potential 
created by the configuration $\varphi (x)$ which, from an appropriate
side, is attractive for both electrons and holes. Then the carriers
activation energy $\Delta$, controlling their concentration
$\rho_n \sim \exp(-\Delta / T)$, plays a role of the $E_{bar}$.

At first sight, the energy $2 E_s$ absorbed by a diverging pair of
solitons is nothing but an upper limit for the dissipation $\Delta E$
for the cross-term relaxation. But actually these two processes have
opposite effects on sliding. Indeed, the relaxation (viewed as a
recombination of the diverging pair) provides a necessary creep over
the defect. But escaping the rtecombination, the divergence of solitons preserves the phase at the
impurity thus preventing the depinning. The subsequent aggregation of
solitons into growing DLs facilitates their annihilation even at large 
distances. Otherwise the proliferation of DLs across the sample
provides the phase slip which occasionally blocks the sliding. 
A detailed analysis is beyond our current discussion.
\vskip .1in
{\bf 2. THE  $I-V$ CHARACTERISTICS.}
\vskip .1in

The state of a bistable impurity depends on its
history. 
As shown at Fig.1, starting with $0< \theta <\pi$ the local state follows the ascending term $E_+$ 
and enters its metastability region $\pi < \theta <2\pi$ where it can fall down to the descending term 
$E_-$ releasing the energy $\Delta E(\theta ) = E_+ (\theta ) -E_-(\theta )$, and complete the circle. 
The tangent $\partial E/\partial \theta$ 
is an instantaneous contribution to the pinning force.
Hence the net pinning force ${\cal F}$ is accumulated from the path interval along  the MS term.
 It reaches the maximum ${\cal E}_{max}$ at $v\rightarrow \infty$ when there is no time for relaxation before the 
MS term is explored along all its allowed length.
 Comparing the dissipated energy $\Delta E n_{i}$ with the one $2e{\cal E}$ gained from the electric
field we obtain $e{\cal E}_{max}= \Delta E_{max}n_{i}/2$ where $n_{i}$ is the linear concentration of strong
impurities. 
We identify it with experimentally observed second threshold field 
${\cal E}_{2} = {\cal E}_{max}$. At finite velocities  $v$ the intrinsic relaxation comes to effect: the MS may 
decay to the stable one by overcoming the barrier even before the end point is reached or a soliton pair  diverges.

 We can write and solve the kinetic equations for the flows of MSs to arrive at our basic expression for the 
pinning force.

 \begin{equation}
 {\cal{F}} =\frac{n_i}{2} \int_\pi^{\theta_e} d\theta {d \Delta E \over d \theta}
 \exp{\left(-\int_\pi^\theta {d \theta_1 \over v \tau (\theta_1)}\right)}
\label{14}
\end{equation}
where $\tau (\theta )$ is the life time along the MS term which is
maximal at $\theta =\pi , \  \tau_\pi =\omega_0^{-1}\exp(E_\pi^b/T)$. 
Here $E_\pi^b=E_{bar}(\pi)$ and  
$\omega_0$ is an attempt frequency.
 Near the end point we find 
 $$\tau \sim \tau_\pi \exp \left( A(\theta_e - \theta)^{3/2}\right) 
 (\theta_e - \theta)^n,\quad A  \sim
 E_\pi^b/T\gg 1, \quad n>1$$
\begin{figure}[thb]
\includegraphics{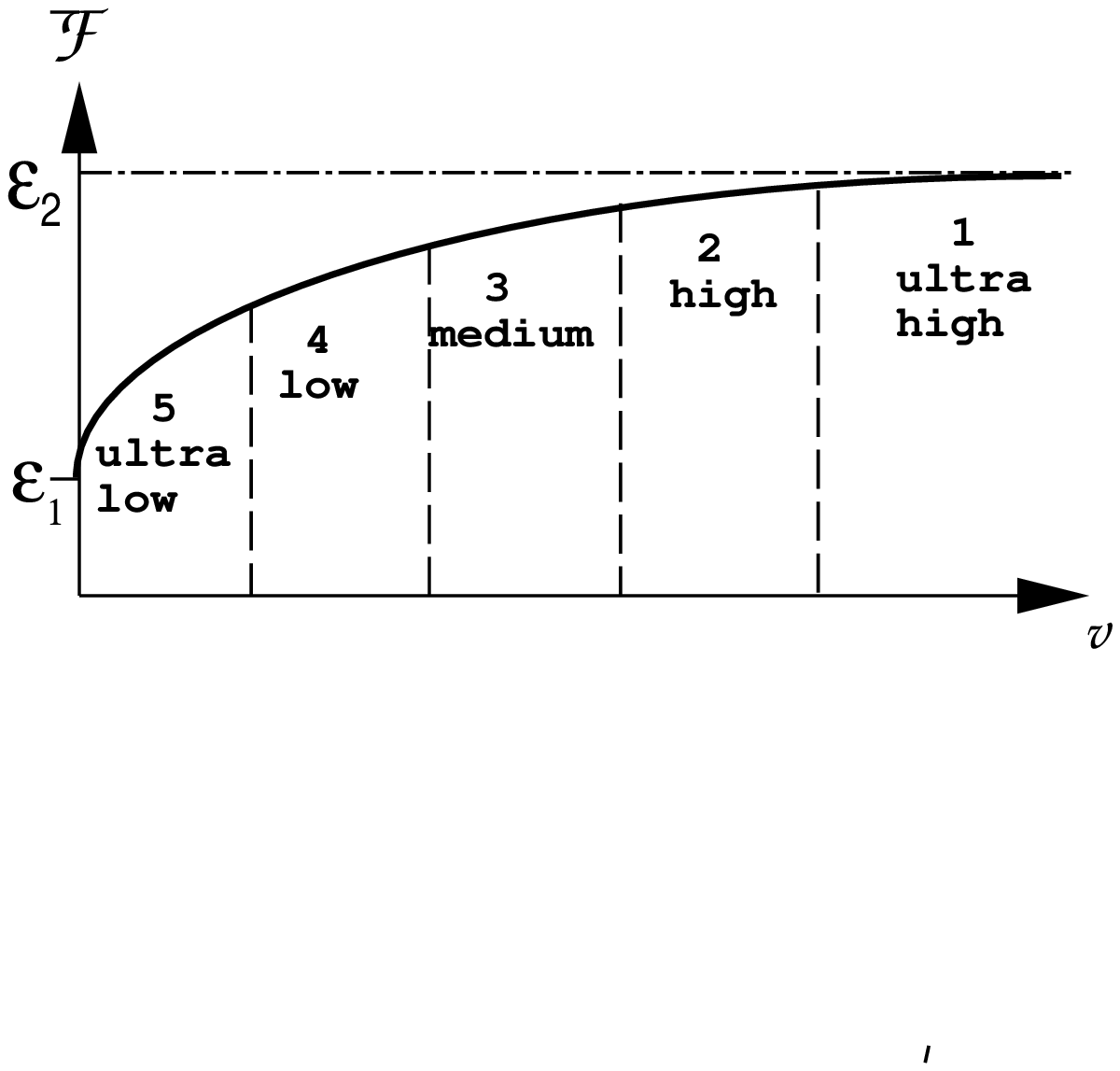}
\epsfxsize 1.5\hsize
\caption
{Various regimes for the MSs contribution to the pinning force. 
A regular viscosity ${\cal F}\sim v$ must be added.}
\end{figure}
The results of calculations fall to several regimes as shown at Fig.2. 
\begin{itemize}
\item{1.} At ultra high $v$  the MS term can reach the close vicinity of $\theta_e$ where the barrier hight is below $T$.\\
$v>\omega _0, \qquad {\cal E}={\cal E}_2 -n_iC(T)v^{-1/(n-1)}$.\\
\item{2.} At high $v$  the end point is approached but the MS  decays at a distance from $\theta _e$ 
where the activation is still required. \\
$\omega_0 >v>v_e \sim (T/\Delta E \tau_\pi ) \exp (\Delta E/T)\ , 
\qquad {\cal E}={\cal E}_2-n_i C(T)[ln(v_0/v)]^{2/3}$\\
\item{3.} At moderate $v$ there is an appreciable advancing along the MS part, $\theta >\pi$, 
of the ascending term but still far from  reaching the end point or completing the full circle. \\
$v_e >v>v_\pi \sim T/\Delta E \tau_\pi \, , 
\qquad  {\cal F}\sim n_i (T/\Delta E^\prime) \ln (v/v_\pi)$  
where $\Delta E^\prime =d\Delta E/d\theta$ at $\theta=\pi$. \\
\item{4.} At low  $v<v_\pi $ the MS decays  as soon as the term becomes
metastable in a vicinity of $\pi $ before any end point is reached. The life
time path  interval is $\delta \theta \sim v\tau _\pi $. Hence at
small $v$ the MSs contribute to the  contribute to the Ohmic  resistance: $%
{\cal E}/v\sim \tau _\pi \Delta E^\prime n_i$. It has an activated behavior via $\tau _\pi $
which may emulate the normal conductivity. \\
\item{5.} At ultra low $v$ the contribution can be seen from some rear regions 
(may be clusters of host imperfections) where barriers  are big enough so that 
$v\tau (E_{bar})\sim 1$ still holds,  hence $E_{bar}\sim T\ln (v_e/v)$ 
At low T the force is given just by their probability 
${\cal P}$.
\end{itemize}

For a natural example 
${\cal P}\sim \exp ( -E_{bar}/T_1)$,  $T_1 = cnst$,
we find that the $I-V$ curve changes from the Ohm low $E \sim v$
at low $v$ to the nonlinear regime $E\sim v^{T/T_{1}}$ with a diverging
differential resistance. This law resembles the dependence of $\varepsilon (T,\omega )$  on $\omega $  
found within the same model \cite{Larkin95}. The above results schematically are present  at the Fig.2.
\vskip .1in  
{\bf{3. DISCUSSION.}}
\vskip .1in

A surprising feature of  the low $T$  sliding regime is the observation of  the two
distinct threshold fields for the onset of the DW sliding \cite{Mihaly88,Itkis90}. Above the first threshold 
${\cal E}_1\sim 10^{-1}\div 1\,V/cm$ the DW starts to slide
collectively while incoherently yet. 
Above the second threshold ${\cal E}_2\sim 1\div 10^2\,V/cm$ 
the sliding turns to be  coherent while  the $I-V$ curve may become very steep.

Since the first threshold  ${\cal E}_1$, withstands very large waiting
times,  it can be only the rest friction due to the
collective pinning. Contrary, in our interpretation,  ${\cal E}_2$ is the high velocity limit of the
dry friction force via the energy dissipation by a moving DW. Above ${\cal E}%
_2$ the power gained from the external field ${\cal E}$ is sufficient to create
necessary MSs to overcome local pinning. Namely at high $v$  the maximal energy $\Delta E$ is
absorbed after each
DW period passing over an impurity which is determined by approaching  the metastability limit  of the 
diverging soliton pair.  
With ${\cal E}_{2}\sim 10V/cm$ and $\Delta E \sim 100K$ we estimate an
average distance between strong defects along the chain as $%
l_{i}=1/n_{i}\sim 10^{-3}cm$. This is a large distance which shows that most of defects are weak.

 The low $v$ sliding conductivity  
$\sigma_s\sim n_i \exp(-E^b _\pi  /T)$ may be related to  an intriguing  appearance  
(in longitudinal conductivity alone!) of a new  activation law  with $E_a<\Delta $ \cite{Nad89}. 
Being attributed earlier to solitonic conductivity, actually this regime may signify on the DW creep through 
strong impurities.  We can also resolve a puzzle which stays unexplained since long time. 
This is the correlation in activation energies for  both normal and collective conductivities.  
(Remind that in linear theory they must be completely decoupled.) 
We achieve it by a very plausible guess that the MSs decay most efficiently by trapping normal electrons 
rather then by the structural fluctuations, then  $\tau \sim
\rho^{-1}_n \sim \exp (\Delta/T)$. 
In any case we obtain

\begin{equation}
\sigma_s = (e/\pi s)(v/E) \sim  (\hbar \omega _p/T_c) ^2(l_i/l_s\tau_\pi)
\end{equation}
(Here $e$ is the electron charge, $s$ is the area per chain,
$\omega_p$ is the parent metal plasma frequency.)
Then for  $\sigma \approx 1 (\Omega \cdot cm)^{-1} \rightarrow 10^{12} s^{-1}$
and taking $l_i$ from ${\cal E}_2$ we estimate $\tau_\pi \sim 10^{-5}s$ 
which is quite comparable with the experimental interval of $v$ which  ranges up to $10^6-10^8Hz$.

Our construction for the $I-V$ curve implies that ${\cal E}_2>{\cal E}_1$ 
which is a quantitative criteria for the predominance of the local pinning. For typical scales of CDWs we have
$
{\cal E}_1 = {\cal E}_{col}\sim E_s l_s n_i^2\rho_n \, , \qquad 
{\cal E}_2={\cal E}_{loc}\sim  E_s n_i
$.
Here the normal carrier density $\rho_n \sim \exp (-\Delta /T)$ appears due to the Coulomb hardening of the DW 
which is limited by a screening by normal carriers.
We find that $E_{loc}$ wins at low $n_i$ and/or low $T$; the crossover from 
$E_{col}$ to $E_{loc}$ takes place at $l_sn_i\sim \rho_n^{-1}\sim\exp (\Delta /T)$. 

\vskip .1in
{\bf 4. CONCLUSIONS}
\vskip .1in

Our approach allows also to describe the Broad Band Noise, the Shapiro steps related to the Narrow Band Noise 
(NBN) and the NBN power spectrum. 
(As well as none of the pinning  models it cannot explain the NBN in linear 
measurements  of $V$ or $I$).
 Our model provides a clue to quantum effects showing that the
 tunneling between the terms destroys the pinning.  
Remind also that the  similar model has been applied earlier to
describe the spectacular frequency dependent peak in temperature
dependence of the dielectric susceptibility 
 \cite{Larkin95}.

Apparently the picture of local pinning is oversimplified and it may be applied only at high enough 
$v$ or $\omega$. Nevertheless it provides some 
insight on  unexplained yet observations  and a guiding (or warning) to theories of much more complicated 
collective processes. 
The adventures come from the explicit treatment of metastable states, their creation and relaxation, 
their relation to plasticity and topological defects. Addressing only to CDWs, 
which feature the most refined experimental information  and the simplest theoretical models, 
we see the following success in explaining the observations. 
First is the distinct  low activation energy for the low $T$ climb  simulating the Ohm law. 
Second, for higher $T$,  is the coincidence of activation energies for  normal and collective conductivities. 
Third is the positive curvature of the $I-V$ curve ($\partial \sigma/\partial E >0$) 
which is frequently observed in a contradiction to prediction of scaling theories. 
The correctly explained curvature and the  effects of the asymptotic saturation of the pinning force 
points towards the effect of the  second threshold field. 

Modern theories of collective pinning exploit the language of MSs in interpretation of their results. 
Nevertheless our unsophisticated  model recovers their potential inconsistencies or misleadings. 
It seems  that the collective treatments are restricted at least to low enough velocities 
where most of the MSs are relaxed. It concerns even the popular approach based on $1/v$ expansions. 
This statement follows from the following observations: \\
1. A fraction of MSs ends up at  the termination points. 
There are those points which determine ${\cal F}(v)$ at high enough $v$ 
but they are not accounted for in perturbational treatments including renormalization group and replica methods.\\
2. Other MS terms do not show this instability which  at first sight  allows for perturbative approaches. 
But it results even in a more obscure effects of  generation of sequence of dislocations. 
Now the $V-I$ dependences are determined by competing of the two processes: 
the annihilation contrary to the aggregation of the DLs. 
While the effects of the DLs on the pinning  has been already noticed \cite{Lee79} 
and estimated (see \cite{Blatter}), 
these processes have not been yet directly accessible except for our simple treatment. 

{\bf Acknowledgements} The authors are indebted to N. Kirova and
B. Shklovskii for help and discussions.

\end{document}